\documentclass[epj-spec]{svjour}
\pdfoutput=1
\usepackage{graphicx,amsmath}
\newcommand{\eeq}{\end{equation}}
\newcommand{\beq}{\begin{equation}}
\newcommand{\ba}{\begin{array}}
\newcommand{\ea}{\end{array}}
\newcommand{\bea}{\begin{eqnarray}}
\newcommand{\eea}{\end{eqnarray}}
\newcommand{\vev}[1]{\langle #1\rangle}
\newcommand{\vp}{\varphi}

\begin{document}
\title{Fundamental constants and their variability in theories of High Energy Physics}
\subtitle{Introductory review for the 2nd.\ ACFC Seminar, Bad Honnef, June 2007} 
\author{Thomas Dent\thanks{\email{t.dent@thphys.uni-heidelberg.de}} 
}
\institute{Institut f{\" u}r Theoretische Physik, Philosophenweg 16 \\
69120 Heidelberg, GERMANY}
\abstract{
The Standard Model of particle physics and the theory of General Relativity (GR) currently provide a good description of almost all phenomena of particle physics and gravitation that have received controlled experimental tests. However, the Standard Model contains many {\em a priori}\/ variable parameters whose values, and whose apparent (near-)constancy, have yet to receive a convincing theoretical explanation. At the same time, GR may now require to be extended or altered at the largest length scales, to account for the recent apparent accelerated cosmological expansion. In this introductory review I present theoretical aspects of the search for explanations of the values and possible variations of fundamental ``constants'', focusing on the possibility of unification of interactions. I also relate cosmological variations to modifications of gravity both locally and cosmologically. }  
\maketitle
\section{Introduction} \label{intro}
\subsection{Motivation: Cosmological variation has occurred!}
It is a fundamental postulate of General Relativity (GR) that the same ``non-gravitational'', local experiment will give the same result at all points in space-time. This is known as the principle of Local Position Invariance (LPI), and forms part of the Einstein Equivalence Principle (EEP).\footnote{See Section~\ref{equiv} for further discussion of relativistic equivalence principles} Yet it is very likely that during the previous evolution of the Universe, elementary particle properties, in particular masses and couplings, were substantially different from their present values. In a hot Big Bang at a temperature above about $1\,$TeV, it is believed that the Higgs field undergoes a phase transition, such that its expectation value becomes zero (as opposed to its present-day value of approximately $247\,$GeV) and thus the leading contribution to most elementary particle masses would vanish. Particle physics under such conditions would be utterly different from today.

This is in one sense an artificial example, since the variation in the expectation value of the Higgs is triggered by temperature, whereas the LPI is supposed to hold after extraneous factors such as temperature are removed. Yet it does show that in the presence of scalar fields that may take different values at different points in spacetime, the assumptions underlying current theories of particle physics and gravitation may become invalid. It is also a warning that properties that are locally valid to high accuracy ({\em i.e.}\/ constancy of particle properties) need not be so when extrapolated over very large distances or cosmological time intervals.

\subsection{Fundamental parameters in particle physics and gravity}
The modern understanding of elementary particle physics and gravitation is encoded in the quantum field theoretical construction of the Standard Model, and in Einstein's theory of GR respectively. The Standard Model has one dimensionful parameter, the Higgs v.e.v.\ $\vev{\phi}$, and on the order of 20 dimensionless parameters which describe the static properties of particles and their interactions via gauge (vector) boson exchange. Gravity introduces one further dimensionful parameter, the Newton constant, which may be exchanged for the Planck mass $M_{\rm P} \equiv (8\pi G_{\rm N})^{-1/2}$ (in ``natural'' units where $\hbar$ and $c$ are unity). Since the value of a dimensionful parameter is only known in terms of units, which themselves have to be defined by some physical system (for example a standard length or mass), the fundamental parameters whose values we can meaningfully discuss include only the {\em ratio}\/ of the Higgs v.e.v.\ to the Planck mass. 

The Standard Model is believed to describe particle interactions from the smallest energies (corresponding to the largest length scales: only electromagnetism can act over macroscopic distances) up to energies at least of order 100$\,$GeV, and possibly much higher. Whether new physics will be needed to supplement or supplant the SM at higher energies is an experimental question that is currently being investigated at the TeVatron at Fermilab and which the forthcoming operation of the LHC at CERN is intended to address directly. If new physics appears then the parameters of the SM may be expressed in terms of those of the new theory: this may lead to an increase, or (hopefully, from the point of view of explanatory power) a decrease in the number of independent parameters. 

One paradigm of how this may happen is {\em unification}\/: the variegation of different particles and interactions may be a symptom of the energy scale we are capable of observing them at. There may be a larger symmetry than we currently observe, which is realized in a nontrivial way or effectively hidden, but would be manifest if we were able to observe interactions at much higher energies. The consequences for fundamental parameters are explained in Section~\ref{unif}: under simple unified models we would expect variations to occur in {\em all}\/ parameters in a correlated fashion. Hence a nonzero variation could test and distinguish between unified models if the observations constrain more than one fundamental parameter.

\subsection{Gravitational and cosmological related effects}
I return to the gravitational sector in Section~\ref{equiv}. The direct physical motivation for considering gravitational effects of spacetime variation is that if the forces that hold matter together depend on position, bodies will be accelerated towards regions where the mass-energy of their internal structure is lower, in addition to the GR gravitational acceleration. This is interesting, and potentially problematic, since these anomalous accelerations break the strong equivalence principle (which tells us that the only gravitational force is due to GR and that the strength of gravity is everywhere constant) and also in general the weak equivalence principle (which tells us that gravitational forces produce the same acceleration of a test body regardless of its composition). 

To get a more quantitative view of equivalence-principle violations, physical models of space-time variation are required. Within a Lorentz-covariant framework this is only possible by introducing a new degree of freedom --- usually a scalar field --- that sources the variation. The choice of model determines the action of this degree of freedom and its couplings to gravity, the Standard Model and possibly other sectors (such as dark matter). Some of the strongest bounds on models of variation then arise from the effects of scalar-mediated forces. Indeed, if a scalar is to source a variation over time scales of the order of the age of the Universe, it should be extremely light and thus mediate very long-range interactions.

Such new degrees of freedom also have nontrivial consequences for the global behaviour of cosmology, since they may contribute nontrivially to the potential and kinetic energy ({\em i.e.}\ stress-energy tensor) at cosmologically large scales. This may either be used as a candidate explanation for the apparent accelerated expansion at recent epochs, as so-called {\em dark energy}, or to bound the possible behaviour of the scalar over cosmic time, in particular to set an observational upper limit on its rate of change.

Given particular models of cosmological variation, one can test them by looking at epochs in cosmology where the variation is expected to be large, or where its effects may be accessible to observation. The earliest time at which bounds can be put on the values of particle physics parameters is that of primordial nucleosynthesis (BBN), which depends on all three fundamental forces (strong, electromagnetic and weak) and on gravity, via the expansion rate of the Universe. Although such bounds are generally at the level of $10^{-2}$--$10^{-3}$ for fractional variations \cite{us07} thus not directly competitive with later spectroscopic results, they apply at a redshift $z\sim 10^{10}$ where the variation might be considerably larger.

\section{The Standard Model and its parameters} \label{SMHiggs}
The definition of the Standard Model comprises the enumeration of matter particles taken to be pointlike fermions, and the description of their interactions via bosons, with the appropriate coupling strengths. These interactions are divided into electromagnetic, weak and strong forces: the essential feature of the Standard Model is that the electromagnetic and weak interactions are both relics of a more fundamental set of forces, technically an SU$(2)\times$U$(1)$ gauge theory, whose effects are masked by the large nonzero value of the Higgs v.e.v., as will be described shortly. 

The strong force also results from a gauge theory, quantum chromodynamics (QCD): its elementary excitations, gluons and quarks, are however not individually observable, being {\em confined}\/ by the strong nonlinear self-interactions of the gluons. The properties of strongly interacting particles (hadrons) are then believed to be explained as ``colour-neutral'' bound states of quarks and gluons; nonperturbative techniques in quantum field theory such as the lattice are required to calculate them. Evidence for individual quarks and gluons only appears at high energies (well above the mass scales of bound states) where, as explained in Section~\ref{RG}, the effective coupling strength becomes weak.

The elementary particle content of the SM can be written via the SU$(3)_C\times$SU$(2)_W\times$U$(1)_Y$ quantum numbers:
\beq \label{SMfermion}
	3\times \left\{ q(3,2)_{1/6} + [ u^c(\bar{3},1)_{-2/3} + d^c(\bar{3},1)_{1/3} ]
	+ l(1,2)_{-1/2} + [ e^c(1,1)_{1} + \nu^c(1,1)_{0} ] \right\}
\eeq
where each representation is written as a left-handed Weyl fermion and we are
including the possible presence of a ``right-handed neutrino'' {\em i.e.}\ a neutral, SU$(2)$ singlet Weyl fermion. The factors of 3 simply denote the three generations, which appear to be identical apart from their mass terms. As required when electromagnetism is a mixture of the SU$(2)$ and U$(1)$ gauge symmetries, the electric charge of the resulting Dirac fermions is derived from a SU$(2)_W$ quantum number (either $+1/2$ or $-1/2$ for a doublet) and the hypercharge $Y$ as
\[
	Q = Y + I_{3W},
\]
where $I_{3W}$ denotes the third component of ``weak isospin''. Hence we can recognize the ``up-'' and ``down-type'' (anti)quarks of charge $\pm 2/3$ and $\mp 1/3$ respectively, and (anti)leptons and neutrinos of charge $\mp 1$ and $0$.

\subsection{Higgs mechanism}
The reason why two weakly coupled gauge groups are needed rather than just the one of electromagnetism is the existence of the weak interactions. The Fermi theory of weak interactions provides an adequate description of low-energy decays. But it is non-renormalizable and has many free parameters, which means that its predictive power is strongly limited; moreover, the theory is actually inconsistent at high energy, since the lowest-order contributions to scattering cross-sections exceed theoretical bounds from unitarity. The introduction of heavy charged {\em vector bosons}\/ $W^{\pm}$ accounts for the weakness and short range of the interaction, and improves the theoretical behaviour of fermion scattering to some extent. But theories with massive vectors also in general have problems with unitarity and nonrenormalizability, and fail to describe ``neutral current'' interactions in which (for instance) a charged lepton and a neutrino scatter without exchanging any charge. 

The solution to these problems was the construction of a renormalizable {\em gauge theory}\/ of electroweak interactions based on the group SU$(2)\times$U$(1)$, which implies four gauge bosons, written as $W_\mu^{\pm}$, $W_\mu^{3}$ and $B_\mu$. The mass of weak vector bosons arises from the Higgs mechanism, {\em i.e.}\ spontaneous breaking of gauge symmetry by a scalar field. Since the Higgs scalar with quantum numbers $(1,2)_{1/2}$ 
\[
	\phi \equiv \begin{pmatrix} \phi^+ \\ \phi^0 \end{pmatrix}
\]
is not invariant under electroweak gauge transformation, if the scalar obtains a nonzero vacuum expectation value (v.e.v.) the gauge symmetry will be hidden. Effectively, the vacuum becomes a medium through which the gauge bosons cannot propagate as massless particles (as also in the theory of superconductivity). The Higgs' hypercharge is such that one component has zero electrical charge: this component may obtain a nonzero value, $\vev{\phi^0}=v/\sqrt{2}$, {\em without}\/ affecting the propagation of the photon, which remains massless.\footnote{In fact, whatever nonzero value of $\vev{\phi}$ we take, one can find one U$(1)$ subgroup that leaves it invariant. A redefinition of fields then brings the situation back to the standard one with electric charge defined as usual.} 

Starting from the non-Abelian gauge theory with a scalar v.e.v.\ one can rewrite the action of gauge bosons in {\em unitary gauge}\/ where the three components of $\phi$ that leave $|\vev{\phi}|$ invariant, the so-called ``Goldstone bosons'', are absorbed by the $W^{\pm}$ and one combination of the $W^3$ and $B$. These vectors obtain masses proportional to $\vev{\phi}$ and to the gauge couplings, while the orthogonal combination of $W^3_\mu$ and $B_\mu$ remains massless: 
\begin{align}
Z_\mu &= \cos \theta_W W_\mu^3 + \sin \theta_W B_\mu \nonumber \\
A_\mu &= - \sin \theta_W W_\mu^3 + \cos \theta_W B_\mu \nonumber  
\end{align}
where
\[
\tan \theta_W = \frac{g'}{g},\ e = g \sin \theta_W,\ M_W=\frac{1}{2}gv,\ M_Z=\frac{M_W}{\cos \theta_W},\ G_{\rm F} = (\sqrt{2}v^2)^{-1}.
\]
Here $g$ and $g'$ are the coupling strengths of the underlying SU$(2)$ and U$(1)$ respectively, while $e$ is the (magnitude of the) electronic charge and $G_{\rm F}$ the Fermi coupling. The components of $\phi$ are accounted for as follows: one excitation about the vacuum value is inherently massive and becomes the {\em Higgs boson}, a neutral massive scalar. Excitations along the other three directions in field space correspond to gauge degrees of freedom along which the potential $V(\phi)$ is constant, and would be massless were it not for the presence of the gauge bosons. In unitary gauge these excitations become the longitudinal modes of the gauge bosons, which thereby obtain three degrees of freedom (helicity $\pm 1$, $0$) rather than the two (helicity $\pm 1$) of a massless vector.

\subsection{Gauge couplings and flavour sector}
Experimentally the parameters of the electroweak theory are found to be $\alpha \simeq 1/137$, $\sin^2 \theta_W\simeq 0.23$, $v \simeq 247$\,GeV, from which the underlying gauge couplings and scalar v.e.v.\ may be deduced. This is a significant simplification over the old Fermi theory, which allowed a large number of undetermined couplings. But we have not yet considered how the Standard Model fermions obtain mass. As written in Eq.~(\ref{SMfermion}) they cannot become massive because two different representations cannot pair up in a gauge invariant manner, except for the singlet neutrino which can be given a Majorana mass. However, the Higgs solves this problem since its quantum numbers allow gauge invariant ``Yukawa couplings'' to be written down
\[
	V_{\rm Yukawa} \sim - h^e_{ij} (\phi\cdot l_{i}) \bar{e}^c_{j} + \mbox{h.c.} + \cdots
\]
where $i$, $j = 1\ldots 3$ run over the fermion family number, and $({\bf a}\cdot{\bf b})$ is an SU$(2)$ invariant product. I have only written the lepton term explicitly. With a nonzero Higgs v.e.v.\ we obtain effective Dirac mass terms 
\[ 
 m^e_{ij} e_{Lj}\bar{e}_{Ri} + \mbox{h.c.} + \ldots
\]
where ${\bf m}^e=\vev{\phi}{\bf h}^e$, the couplings $h$ and masses $m$ being 3-by-3 matrices in family space.

Considering the up- and down-type quarks and charged leptons, one could {\em a priori}\/ have 27 complex couplings, greatly increasing the number of parameters.\footnote{The neutrino sector introduces still more, in general both Dirac and Majorana masses.} Most of these parameters are not physical: by field redefinitions under a global U$(3)^5$ flavour symmetry which acts on each Weyl fermion representation separately, one can bring the charged lepton mass matrix and one of the quark mass matrices (say ${\bf m}^u$) to diagonal real form. The remaining quark mass matrix will still in general be complex and non-diagonal. 

Then a further field redefinition which treats the up- and down-type parts of the SU$(2)$ quark doublet differently, and thus is not a symmetry of the weak gauge interaction, brings all matrices into real diagonal form. This is done at the cost of introducing a 3-by-3 unitary mixing matrix into the charged current ($W^{\pm}$) interactions between quarks, the {\em Cabbibo-Kobayashi-Maskawa (CKM) matrix}. This has 4 parameters: the 9 parameters of a general unitary matrix, reduced by 5 through redefining the Dirac fermions by complex phases.

Then the flavour sector contributes the following parameters: three charged lepton masses $m_e$, $m_\mu$, $m_\tau$; down-quark masses $m_d$, $m_s$, $m_b$; up-quark masses $m_u$, $m_c$, $m_t$; and the 3 quark mixing angles $\theta_{1,2,3}$ and one complex phase $\delta_{\rm CKM}$ of the CKM matrix.\footnote{Nonzero neutrino masses typically contribute at least three more masses, three mixing angles and one complex phase, however this count is model-dependent.} 

At this point we have defined the Standard Model so far as it can be treated by perturbative quantum field theory with weak coupling strength and at lowest order. Higher order corrections do not change the basic picture of electroweak theory. However, the strong interactions cannot be described in this way. In particular, at low energies they are dominated by nonperturbative effects that introduce new mass scales, in addition to the Higgs v.e.v.\ and SM fermion masses. Also, in order to consider more fundamental theories that could lead to functional relations between the SM parameters, we will have to calculate with exponentially large ratios of mass or energy scales, for which lowest-order calculations are completely inadequate. Both these problems will be addressed in the next section.

\section{Energy scale dependence of couplings and unification} \label{unif}

\subsection{Loop diagrams and RG evolution} \label{RG}
In quantum field theory (QFT) calculations of scattering processes and bound state energies, perturbative contributions are ordered by powers of $\hbar$. Higher order contributions include higher powers of coupling constants which we denote as $g$, and some integrals over four-momenta of internal lines in Feynman diagrams, representing virtual fields which do not appear in the initial or final state. Such integrals can be schematically written as
\[
 \lim_{p_{\rm max}\rightarrow \infty} \int_0^{p_{\rm max}} dp\, f(s,t,u,m,p)
\]
where $s$, $t$, $u$ are the Mandelstam variables (certain finite combinations of initial- and final-state four-momenta), $m$ is a particle mass and $f$ is some known algebraic function. Depending on the theory, the integrals may be finite, or may diverge as a logarithm or power of the upper limit of integration. Thus in general the integral must be regularized, or cut off. This occurs either by a mathematical procedure in which an arbitrary energy scale $\Lambda$ is introduced by hand; or by a {\em physical cut-off}\/\! where the structure of the theory changes at a given, large energy scale $M_{\rm uv}$ and it is not appropriate to consider point-like virtual particles above this energy. The virtual particles then have some tightly-bound substructure which becomes evident at a length scale $1/M_{\rm uv}$; below this length one should start to consider the individual constituents as propagating fields instead. 

The first possibility, regularization by hand, can only be useful if the physical quantities that emerge from summing contributions at different powers of $\hbar$ can be made independent of the arbitrary regularization scale to sufficiently good precision. Remarkably, in a class of QFTs including the Standard Model, this is the case: such theories are called {\em renormalizable} \cite{Weinberg}. A renormalized coupling $g_R$ may be defined, which has a functional dependence on $\Lambda$ such that measurable quantities calculated to any given order in $g_R$ no longer depend on the regularization. 

However, the result may include logarithms of $E/m$, where $E$ is a typical energy scale of the process (controlling the magnitude of $s$, $t$, $u$). In order for perturbation theory to work, the size of terms should decrease sufficiently quickly as one goes to higher orders of the expansion in $\hbar$ and $g^2$: hence the presence of large logarithms $\ln (E/m)$ may be problematic. 

The appropriate solution is {\em renormalization group evolution}\/ of coupling constants. Essentially we define a renormalized coupling $g_\mu$ depending on an arbitrary energy scale $\bar{\mu}$ which is allowed to vary.\footnote{The bar distinguishes it from the proton-electron mass ratio $\mu$.} In perturbation theory we may then calculate the value of $g_{\mu'}$ at a different scale to any given order as long as $\ln(\bar{\mu}'/\bar{\mu})$ is small. Now taking $\ln(\bar{\mu}'/\bar{\mu})\rightarrow 0$ a differential equation is obtained
\beq \label{betafn}
	\frac{d}{d\, \ln \bar{\mu}} g_\mu = \beta(g_\mu) 
\eeq
where the ``beta-function'' on the RHS may be expanded in powers of $g_\mu$ and will depend on the gauge group and matter content of the theory. For some process of interest occurring at an energy or momentum scale $\bar{\mu}$, the size of loop corrections is minimized, and thus perturbation theory is most accurate, by expanding in powers of $g_\mu$: this value is obtained from the value $g_\mu'$ at another scale (characterizing some other physical process) via the RG equation (\ref{betafn}). 

In QCD (SU$(3)$ Yang-Mills theory with fundamental Dirac fermion representations) we have at first order
\beq \label{betastrong}
	\beta(g_s) = -\frac{g_s^3}{4\pi^2}\left(\frac{11}{4} - \frac{n_f}{6}\right)
\eeq
where $n_f$ is the number of quark flavours contributing, {\em i.e.}\ quarks with mass $m_q \leq \bar{\mu}$.\footnote{Particles heavier than the RG scale contribute negligibly as their propagators are suppressed by powers of mass.} The negative sign of $\beta(g)$ has two notable consequences. First, at large momentum transfer the coupling decreases monotonically and loop corrections become less and less significant. QCD is {\em asymptotically free}\/ and perturbative at high energies: experimentally, the ``running fine structure constant'' $\alpha_s(M_Z) \equiv g_s(M_Z)^2/4\pi \simeq 0.118$. 
\begin{figure} \label{QCDrun}
\centerline{
\resizebox{0.5\columnwidth}{!}{ \includegraphics{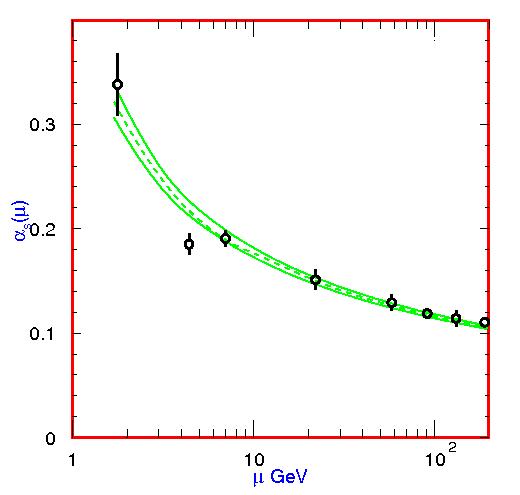} }
	}
\caption{Experimental running of $\alpha_s\equiv g_2^2/4\pi $ with energy (PDG review, 2006)}
\end{figure}
Second, on integrating Eq.~\ref{betafn} a constant $\Lambda_c$ with dimensions of energy enters: the solution of the one-loop RG equation is
\beq \label{alphaslambda} 
	\alpha_s(\bar{\mu}) = \frac{12\pi}{(33-2n_f)\ln(\bar{\mu}^2/\Lambda_c^2)}
\eeq
and we see that $\alpha_s$ diverges as the energy scale approaches $\Lambda_c$ from above. The coupling becomes strong, higher-order terms should be considered and non-perturbative effects eventually dominate, such as confinement and the formation of a chiral quark condensate, which cannot be described within this framework: the degrees of freedom at low energies are no longer gluons and quarks.

Nevertheless the behaviour of QCD (up to the effects of quark masses) is determined completely by specifying the value of $\Lambda_c$, and all energy scales associated with strong coupling effects such as the formation of hadrons are then proportional to $\Lambda_c$. Thus we obtain some handle on the functional dependence of strong-interaction physics on the value of $\alpha_s$ defined at some high energy scale. The value of $\Lambda_c$ (defined in some particular renormalization scheme) is then found by comparing perturbative calculations to scattering data and is of order $250\,$MeV.

The SU$(2)_W$ and U$(1)_Y$ gauge couplings of the Standard Model also show scale dependence:
\begin{eqnarray}
	\beta(g) &=& -\frac{g^3}{4\pi^2}\left(\frac{11}{6} - \frac{n_g}{3}\right) \nonumber \\
	\beta(g')&=& -\frac{g'^3}{4\pi^2}\left( - \frac{5n_g}{9}\right) \nonumber
\end{eqnarray}
where $n_g$ counts generations of fermions. We may derive the running electromagnetic coupling $\alpha_{\rm em}(\bar{\mu})$ and weak mixing angle $\theta_W(\bar{\mu})$ (the quoted value $0.23$ applies for $\bar{\mu}=M_Z$). However, the fine structure constant $\alpha$ is defined in the limit of zero momentum transfer (indeed at momenta lower than $m_e c$ the beta-function vanishes) therefore possible cosmological variations of $\alpha$ are independent of the question of RG scale dependence. 

\subsection{Gauge unification and SUSY}
It was noticed some time ago that the weakest Standard Model gauge coupling becomes stronger with increasing energy scale, while the strong coupling becomes weaker; one may naturally ask whether at some high energy $M_X$ the three forces could have a single common coupling strength and a common ``unified'' identity as aspects of a single more fundamental interaction. 

The Standard Model already contains examples of broken or hidden symmetries. The $W^{\pm}$ and $Z^0$ bosons result from a partially broken gauge symmetry in the Higgs phase, where the nontrivial properties of the vacuum state hide the gauge invariance. The strong interaction, meanwhile, exhibits {\em confinement}, meaning that particles charged under the QCD gauge symmetry can never be observed: the energy cost of isolating a colour-charged particle would be larger than that of creating new particles to neutralize its charge. These two possibilities may also occur in new physics: there may be a larger gauge symmetry that is spontaneously broken, or the known particles may actually be composite bound states of some theory formulated at higher scale. I will focus on the former possibility simply because it is easier to calculate the consequences, as the coupling strength of such a theory turns out to be small and nonperturbative effects do not directly contribute.

The first step towards unification concerned the matter representations of the SM. It was proposed \cite{PatiSalam} that the charged lepton SU$(2)_W$ doublets and singlets should be unified with the quark doublets and singlets in fundamental representations of SU$(4)$. Moreover, an SU$(2)_R$ symmetry, also broken at high scale, unites the up-(neutrino-)type and down-(lepton-)type weak-singlet fermions. At leading order such {\em Pati-Salam}\/ unification enforces relations between the masses of elementary fermions. The most convincing example is the heavy fermions of the third generation \cite{AllanachQuadruple}, for which one obtains at the unification scale 
\[
	m_\tau \tan \beta = m_b \tan \beta = m_t = m_{\nu_\tau},\qquad \bar{\mu} = M_X
\]
up to small corrections, where $m_{\nu_\tau}$ is the Dirac mass of the tau neutrino and $\tan \beta$ is the ratio of the v.e.v.'s of two electroweak Higgs doublets required by the theory. The observed difference between $\tau$ and $b$ quark masses is then accounted for by different RG running down from the unification scale, where the interactions of the quarks and leptons split off from one another.

Most attention has focused on models that allow the fundamental forces to be unified (GUTs). The simplest mechanism for gauge unification is analogous to the symmetry-breaking of the Standard Model, but with a larger gauge group, in which some massive scalars receive nontrivial v.e.v.'s and their Goldstone modes are ``eaten'' by gauge bosons, leaving only the SM gauge bosons massless. Common symmetry breaking schemes start with a unified group SU$(5)$ or SO$(10)$. 

There are two notable consequences for the SM parameters. First, there may be algebraic relations between the Yukawa couplings and hence the masses and mixings of the SM fermions, as in Pati-Salam unification. Such relations depend strongly on the sector responsible for gauge symmetry breaking. Second and more importantly for us, the effective ``fine structure constants'' of the SM groups at the unified scale $M_X$ obey a fixed relation
\[
	g_s^2(M_X) = g^2(M_X) = g_1^2(M_X) = g_X^2(M_X),\qquad g_1^2=\frac{5}{3}g'^2.
\]
Note here that while the coupling strength of non-Abelian groups arise directly from the self-interaction of vector bosons, the coupling strength of U$(1)$ is known only up to a constant factor which also scales the charges of matter fields. One needs to identify the charges of unified matter representations under the $U(1)$ subgroup with the known SM charges to obtain a correct ``hypercharge normalization'', resulting in the factor $5/3$ which occurs for many (though not all) unified models. The experimentally known coupling strengths at $\bar{\mu}=M_Z$ can be extrapolated to higher energies to test the unification relation. 
\begin{figure}
\centerline{
	\includegraphics[width=5cm, viewport = 155 505 445 720, clip]{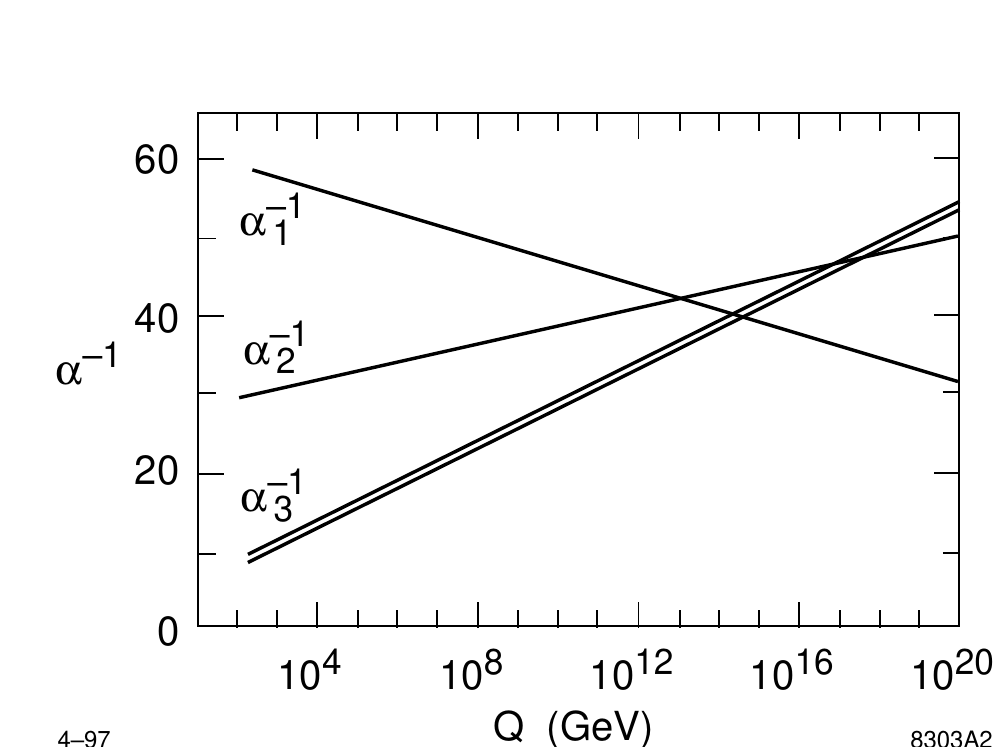}
	\includegraphics[width=5cm, viewport = 155 505 445 720, clip]{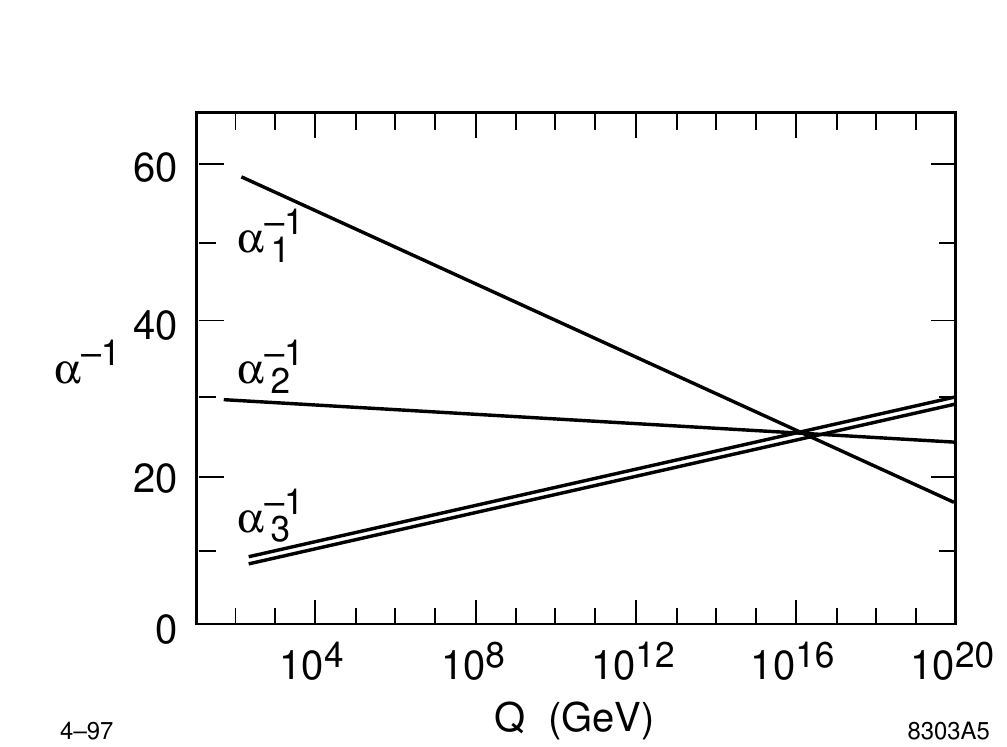}  
	}
\label{unifig}
\caption {RG dependence of gauge couplings in SM and SUSY unification (M.~Peskin) } 
\end{figure}

The result (Fig.~\ref{unifig}, left-hand graph) is not very accurate for the simplest unification models.\footnote{Such models are ruled out also by predicting too rapid proton decay.} Non-minimal models including extra, heavy matter multiplets or multiple stages of symmetry-breaking can repair the discrepancy, but lose predictivity for our purpose unless additional assumptions are made. 

However, another way of adding extra particles has the effect both of fixing grand unification unambiguously and improving other properties of unified theories. This is {\em supersymmetry}\/ (SUSY), a symmetry that relates each fermion in a theory to a boson with the same charge and mass (and vice versa). Thus the total count of particles is doubled.\footnote{It turns out not to be feasible for any SM fermion to be the superpartner of a SM boson.} The particle physics motivation for introducing SUSY is the technical naturalness of the mass parameter in the Higgs potential $V(\phi)$. This was omitted from previous discussion since the parameters of low-energy physics can be found without knowing the exact shape of $V(\phi)$. 

The scalar mass is the only quantity in the SM whose value receives quantum corrections that diverge {\em quadratically}\/ with the cutoff. Considering a more fundamental theory defined at a scale $M_X$, successive contributions to the scalar mass-squared in the low-energy theory are of order $M_X^2$ times some factors of coupling strengths. But the Higgs mechanism cannot work as observed unless the mass parameter is of order $(100\,$GeV$)^2$. Thus the scalar mass in the fundamental theory at $M_X$ must be fine-tuned to extremely high accuracy: moreover, the fine-tuning must account for the {\em exact}\/ quantum corrections, which cannot be found in perturbation theory with sufficient accuracy. However, precisely these quantum contributions to scalar mass vanish within supersymmetry, due to cancellation between fermions and bosons.

Since the particle content is altered by SUSY, so are the RG equations, and it turns out that the fit of the measured SM couplings with a single grand unified coupling is much improved (Fig.~\ref{unifig}, right-hand graph) with a unification scale $M_X\simeq 2\times 10^{16}\,$GeV and a unified coupling $\alpha_X \equiv g_X(M_X)^2/4\pi \simeq 1/24$.\footnote{The unification was probed at LEP years {\em after}\/ the first SUSY-GUT models were constructed.} Now we do not observe superpartners with the same mass and charge as the SM particles: however, it is possible to break supersymmetry but still preserve both the technical naturalness of the Higgs sector and grand unification. {\em Soft SUSY-breaking}\/ introduces masses of order $100\,$GeV$\,\sim\vev{\phi}$ for the superpartners. Intuitively, this change should not affect properties that depend on the very large ratio $M_X/\vev{\phi}$. We will consider the possible relation between superpartner masses and the Higgs v.e.v.\ in the next section.

Now we may invert Eq.~(\ref{alphaslambda}) with $n_f=3$ and differentiate to obtain the dependence of $\Lambda_c$ (and thus the mass scale of hadrons) on the unified coupling:
\beq \label{varLambda}
	\Delta \ln \frac{\Lambda_c}{M_X} = \frac{2\pi}{9\alpha_X}\Delta \ln \alpha_X 
	+ \frac{2}{27}\Delta \ln \frac{m_cm_bm_t}{M_X^3} 
	+ \frac{2}{9}\Delta \ln \frac{m_{\tilde{q}}m_{\tilde{g}}}{M_X^2} 
\eeq
where the last two {\em threshold}\/ terms on the RHS arise from the change in the beta-function when the RG scale $\bar{\mu}$ passes through the mass of a charged particle. Here $m_{\tilde{q}}$, $m_{\tilde{g}}$ are the masses of the quark and gluon superpartners. If we similarly calculate the effect of RG running from $M_X$ down to zero momentum on the electromagnetic coupling we find
\beq \label{varalpha}
	\Delta \ln \alpha = \frac{8\alpha}{3\alpha_X} \Delta \ln \alpha_X 
	+ \alpha\sum_i \frac{Q_i^2f_i}{2\pi} \Delta \ln \frac{m_i}{M_X}
\eeq
where the threshold term runs over all electrically charged fields with masses $m_i$. Note that in both cases the threshold terms are higher order in $\alpha$ or $\alpha_X$. Hence formally to leading order, the proton mass variation is expected to be \cite{CalmetFritzsch} 
\beq \label{mpvar1}
	\Delta \ln \frac{m_p}{M_X} \simeq \frac{\pi}{12\alpha} \Delta \ln \alpha \simeq 36 \Delta \ln \alpha. 
\eeq
For instance, if there is a nonzero variation of $5\times 10^{-6}$ in $\alpha$ at some epoch, the corresponding variation of $m_p/M_X$ is about $1.8\times 10^{-4}$, which lies well outside observational bounds on $\mu\equiv m_p/m_e$ variation from spectroscopy \cite{mubounds}  unless the variation of the electron mass $\Delta \ln (m_e/m_X) = (35\pm 5) \Delta \ln \alpha$. We will however see that accounting for possible variations of SM fermion and superpartner masses can radically change the relation between variations of $\mu$ and $\alpha$.

\subsection{Hierarchy problem and varying mass scales}
So far the origins of the Higgs v.e.v.\ $\vev{\phi}/M_X\simeq 10^{-14}$ and small Yukawa couplings ({\em e.g.}\ $m_d/m_b \simeq 1/700$) have not been discussed. The first of these small ratios is known as the {\em hierarchy problem}. As already discussed, SUSY stabilizes the hierarchy by ensuring the Higgs potential receives only logarithmic corrections proportional to the superpartner masses. An alternative route is for the Higgs to be a composite state of a theory that becomes strongly coupled around the electroweak scale, $\Lambda_{TC}\sim 1\,$TeV, analogously to the strong coupling regime of QCD. In such {\em technicolor}\/ models the theory above $\Lambda_{TC}$ no longer includes the Higgs and, as a weakly-coupled gauge theory, has only a logarithmic dependence on physics at higher energy scales.

Then the ratio $\vev{\phi}/M_X$ will depend exponentially on the technicolor gauge coupling at high energies:
\beq \label{ewsbexpo}
	\frac{\vev{\phi}}{M_X} = k \alpha_{h}^n e^{-2\pi m/b_{h}\alpha_{h}(M_X)}
\eeq
where $b_h$ is the first term in the expansion of the beta-function (Eq.~(\ref{betafn})) with a factor $-g^3/16\pi^2$ removed: $b_{h}>0$ means asymptotically free, {\em cf.}\ Eq.~(\ref{betastrong}). Then if $k$ and $n$ are order 1, we may estimate $m/b_{h}\alpha_{h}(M_X)\simeq 32$ independently of the exact model used \cite{LangackerSS,me03} given that $M_X$ is order $10^{16}\,$GeV. Hence $\Delta \ln \vev{\phi}/M_X \simeq 32 \Delta \ln\alpha_h(M_X)$.

Within SUSY, essentially similar models are considered for a ``hidden sector'' where SUSY is broken at a scale exponentially below $M_X$, in order to produce superpartner masses $\tilde{m}$ of order the electroweak scale. Since quantum contributions to the Higgs potential are comparable to the superpartner masses, $\tilde{m}$ should not be much larger than $\vev{\phi}$ except in special cases. In many unified models electroweak symmetry-breaking is caused exactly by such ''radiative corrections'' and one may thus set $\Delta \ln \vev{\phi} \simeq \Delta \ln \tilde{m}$. The relation of Eq.~(\ref{ewsbexpo}) applies here also where $\alpha_h$ now applies to the SUSY-breaking hidden sector. If the hidden sector or technicolor group is then unified at the GUT scale we obtain a relation between the variations of the Higgs v.e.v.\ and superpartner masses, and that of the unified coupling. 

However, the scale of electroweak symmetry-breaking may be strongly dependent on details of the model, including the variation of the top Yukawa coupling \cite{CampbellO} which is {\em a priori}\/ unknown. Hence we will keep the variation of $\vev{\phi}$ and $\tilde{m}$ as adjustable parameters:
\beq \label{betadefin}
	\beta_v \equiv \frac{\Delta \ln(\vev{\phi}/M_X)}{\Delta \ln \alpha_X},\
	\beta_S \equiv \frac{\Delta \ln(\tilde{m}/M_X)}{\Delta \ln \alpha_X},
\eeq
bearing in mind that they are likely to be of order 30 in the simplest unification scenarios. 
We can insert these 
variations in the Higgs v.e.v.\ and superpartner masses into the variations of $\Lambda_c$ and $\alpha$, Eqs.~(\ref{varLambda}-\ref{varalpha}), to obtain \cite{me03}
\beq
	R \equiv \frac{\Delta \ln \mu }{\Delta \ln\alpha } \simeq \frac{0.54/\alpha_X -0.6\beta_v +0.35\beta_S}{0.022/\alpha_X + 0.002\beta_v +0.011\beta_S }
\eeq
where for non-supersymmetric theories we set $\beta_S=0$. Here we have also taken account of the small contributions of light quark ($u$, $d$ and $s$) masses to the proton mass, which mean that $m_p$ is not exactly proportional to $\Lambda_c$. 

Now for $\beta_v = \beta_S=0$, we obtain $R\simeq 25$, independent of $\alpha_X$. This is a smaller variation of $\mu$ than Eq.~(\ref{mpvar1}) due to the more accurate treatment of the proton mass, but still observationally problematic. 

For $\beta_v \simeq 32$, $\beta_S=0$, and the choice $\alpha_X\simeq 1/24$, we obtain $R\simeq -11$ with an uncertainty of a few. With $\alpha_X\simeq 1/42$, which may correspond more closely to non-SUSY unification, we have $R \simeq 3.5$ with an uncertainty of about $5$.

In the case of varying superpartner masses, $\beta_v = \beta_S \simeq 32$ and $\alpha_X\simeq 1/24$, we find $R\simeq 5\pm 4$. Hence the relation between varying $\mu$ and $\alpha$ is strongly sensitive to possible variations in the hierarchy of the Higgs v.e.v.\ or superpartner masses relative to $M_X$. Only a relatively small range of values of $\beta_v$ and $\beta_S$ is compatible with a nonzero variation of $\mu$ being the same size as a variation of $\alpha$.

\subsection{String theory}
The relation between particle theory and string theory is complicated and cannot be summarized briefly. I will only make a few general remarks on the status of coupling strengths and mass scales in string models. Contrary to the field theory of point particles, string theory has one dimensionful fundamental parameter $\alpha'$, the inverse string tension ({\em i.e.}\ energy per unit length) of dimension $[E]^{-2}$. One can equivalently define the ``string scale'' $M_S\sim 1/\sqrt{\alpha'}$. 

At energies much below $M_S$, only the lowest energy string excitations contribute to the dynamics: these are expected to behave like point particles to good approximation, and one can construct an {\em effective field theory}\/ which describes their interactions in a flat space-time background, by matching field theory scattering amplitudes to string theory amplitudes. This is the starting point for much string model-building. One notable feature at this level is that the string coupling constant---the amplitude for one string to split into two---is given by the value of an effective scalar field, the dilaton. Thus we would need to know the dilaton potential, or at least its minimum, to make any statement about couplings.

Supersymmetric string theory is, furthermore, necessarily defined in 10-dimensional spacetime. Hence it is necessary to make six spatial dimensions invisible to low-energy excitations. This is achieved by {\em compactification}: a six-dimensional space is found which, in conjunction with 4-d spacetime, satisfies the equations of motion, and only modes with zero momentum along the six extra dimensions have sufficiently small mass to be experimentally accessible.\footnote{This means that the size of compact extra dimensions should be at least as small as TeV$^{-1}$, except in a subset of ``brane world'' models where the SM cannot propagate along some direction.}

The details of four-dimensional physics then depend strongly on the six-dimensional space (and any fields that have nonzero components along it). In many cases there are continuous families of such spaces, parameterised by degrees of freedom that correspond to more scalar fields in 4-d spacetime, the {\em moduli}. Then we are interested in the minima of the potential as a function of the dilaton and the moduli; or if there is no minimum (as for an exponential potential \cite{Piazza01}), the speed of possible cosmological evolution of the fields. Recent developments in string model building have shown that most such scalars can become massive, {\em i.e.}\ the potential has minima in these directions. But the stabilization of all moduli and their masses and couplings to SM fields is an open question. We will see in the next section that the couplings of any light, cosmologically evolving scalar to the SM must be very weak (even weaker than gravity); this seems unlikely to be satisfied in most string models.

In principle in a given string model, once the potential of the moduli is known, algebraic relations between the SM gauge couplings and Yukawa couplings should follow. In practice no model combines (semi-)realistic gauge symmetry matter and content with stabilized moduli to the extent of making predictions. Most ``semi-realistic'' models resemble grand unified theories in that there are simple relations between gauge couplings at the string scale, which however generally depend on the values of moduli. The details of how the SM gauge groups emerge and how the Yukawa couplings are determined are, however, different in string models. The main advance of strings over GUTs is that string models explicitly include gravity: the ratio $M_S/M_{\rm P}$ (the strength of gravity) is predicted as a function of the dilaton and moduli, leading to an extra condition on the values they may take consistently with observation.

\section{Deviations from GR and cosmological variations} \label{equiv}

In particle theory we are required, due to the many precise tests of Lorentz invariance in microscopic physics \cite{MattinglyLRR} to start from an action that is Lorentz covariant. Each of the fields representing a particle is a representation of the Lorentz group and the fields may only be combined in certain ways. This ensures that physical results will be unaltered under rotations, translations and boosts, after having controlled for environmental factors that do not respect this symmetry. Conservation laws such as energy and momentum are thus also built in. 

A situation where fundamental parameters are {\em not}\/ unchanged under translation in time or space is then at first sight problematic: what becomes of the symmetry and conservation principles that give physics much of its simplicity and predictivity? For instance, merely introducing (in QED) a fine structure constant $\alpha({\bf x})$ that depends on space-time position will break energy and momentum conservation and gauge invariance, and removes any predictivity from the theory, since the function $\alpha({\bf x})$ is {\em a priori}\/ completely undetermined. 

Instead, we replace the unknown function by a dynamical field, a new degree of freedom whose behaviour is determined by its own action and couplings to the rest of the theory. We have already seen examples of this in the Higgs mechanism, where the SM fermion masses are functions of the Higgs v.e.v., and in string theory, where every coupling strength is due to some background field taking a particular value. Considering the action as a whole, Lorentz covariance is still valid: the variation in measurable quantities arises because the {\em solution}\/ for the extra degree of freedom has a nontrivial space-time dependence. I will refer to such a field as a {\em cosmon}\/ and denote it generically as $\varphi$. (It is assumed to be a scalar since other types of field generally cause gross violations of Lorentz symmetry if given a vacuum value.) An apparent violation of energy conservation in the SM, for example, would actually be a transfer of energy to or from $\vp$, and should be correlated with other (in principle) measurable effects. 

Our task is then to relate the action of $\vp$ to observable effects. Although the action contains {\em a priori}\/ unknown functions of $\vp$, it may have predictive power in that the same configuration of the scalar should always produce the same observable effect: we assume that observable quantities are smooth and well-behaved functions of $\vp$.

\subsection{Equivalence principles and their violation}
Considering variations with respect to gravitational as well as microscopic physics, we have to allow for the observational success of GR \cite{WillLRR,GRprocs} on scales from the laboratory to beyond the Solar System, which strongly restricts the theories we may write down. Note again that the appearance of space-time dependent fundamental parameters violates most of the basic principles of the theory. Consider first the {\it Einstein equivalence principle}\/ (EEP): it consists of three parts.
\begin{enumerate}
\item Weak equivalence principle (WEP): The trajectory of a freely falling test body is independent of its structure and composition.
\item Local Lorentz invariance (LLI): The outcomes of non-gravitational experiments are independent of the velocity of their (freely-falling) reference frame.
\item Local position invariance (LPI): Outcomes of non-gravitational experiments are independent of their position in space-time.
\end{enumerate}
The mass-energy of a test body depends on the fundamental parameters, and it is found  (Nordtvedt) that in any situation where these are varying over spacetime the body will undergo an acceleration in addition to the gravitational acceleration due to GR. In general the masses of different test bodies will have different dependences, thus WEP is violated. The gradient of any given parameter defines a direction in spacetime, thus the boost and rotation invariance implied in LLI is broken, as (obviously) is LPI. 

The {\em strong equivalence principle}\/ (SEP) is analogous to EEP, except that the universality of free fall holds also bodies with non-negligible gravitational self-energy, and the invariance principles apply also to local gravitational experiments. Violation of SEP occurs specifically if $G_{\rm N}$ varies (in units where particle masses are constant), for example in Brans-Dicke theory. Then the acceleration of a falling body will have a contribution from the gradient of its gravitational self-energy, which will depend on its mass and shape. The results of gravitational experiments (but not necessarily others) are also clearly dependent on space-time position and velocity in this case. 

The anomalous accelerations that signal non-GR effects may be simply expressed in terms of the gradients of the mass-energy of freely falling particles:
\beq \label{delta_a}
	\delta \overset{\rightarrow}{a} = 
	-\overset{\rightarrow}{\nabla} \ln \frac{M}{M_{\rm P}} =
	- \frac{\partial \ln (M/M_{\rm P})}{\partial \ln G_k}
	\overset{\rightarrow}{\nabla} \ln G_k
\eeq
where $M$ is the total mass-energy including binding energy and self-energy, and $G_k$ are the values of fundamental parameters on which $M$ may depend. In what follows we will neglect the gravitational self-energy, since this component is negligibly small for bodies whose accelerations have been most accurately measured. 

Already at this level one can compare two methods of obtaining limits on variations of $\alpha$ within the Solar System. First via the dependence of atomic clock frequencies \cite{Fortier07} as the Earth moves through the gravitational potential in its elliptical orbit; or second, in an E{\" o}tv{\" o}s type experiment, via the differential acceleration $\eta \equiv |\delta a_1 - \delta a_2|/|g|$ of test bodies with different fractions of electromagnetic binding energy $B_{\rm em}/M\equiv -\partial \ln M/\partial \alpha$. We have (up to a sign)
\beq
	\eta = \frac{\partial \ln (M_1/M_2)}{\partial \ln \alpha} \frac{|\nabla \ln \alpha|}{g}.
\eeq
Note that the last factor on the RHS is the ratio of the gradient in $\alpha$ to that in the Newtonian potential. Now despite recent improvements in atomic clock techniques, the E{\" o}tv{\" o}s method is currently far more sensitive to spatial variations \cite{Nordtvedt02}. The limits on $\eta$ for various pairs of test bodies are at the $10^{-13}$ level \cite{Baessler&Schlamminger} with $\Delta_{12}(B_{\rm em}/M)$ of order $10^{-3}$--$10^{-4}$. For the elliptical orbit of the Earth, the variation in the gravitational potential is a few times $10^{-10}$, hence the seasonal variation in $\alpha$ consistent with bounds on $\eta$ is at most of order $10^{-19}$, barring some unusual cancellation in Eq.~(\ref{delta_a}) \cite{Shawseasonal}. Given that the clock frequencies vary as $\omega_i \propto \alpha^n_i$ with $n_i$ of order 1, frequency stability at the $10^{-19}$ level would be required for the clock method to be competitive. Alternatively clocks could be sent into deep space to take advantage of a much greater variation in the  

The absolute size of anomalous accelerations $\delta a$ is also limited by tests of planetary motion and the form of the metric within the Solar System. This bound is relevant to Brans-Dicke theory where all test bodies experience the same acceleration due to the scalar interaction: the inverse scalar coupling strength $\omega$ in the theory is limited to be greater than $40,000$ \cite{Cassini}. However, in theories where particle physics parameters such as $\alpha$ and $\mu$ vary, differential accelerations give a much stronger bound.

\subsection{WEP violation via scalar couplings}
To relate local bounds on $\eta$ to more fundamental theories and to cosmological models we should write an action to describe the dynamics of the scalar $\bar{\varphi}\equiv \varphi/M_{\rm P}$ (here made dimensionless) and its couplings. By means of metric and field redefinitions this can be put into a form with the standard Einstein-Hilbert action for gravity and a canonical kinetic term for the scalar: the only remaining freedom is in the coupling functions of the scalar to SM matter and gauge fields. We may write a {\em low energy effective action}\/ (strictly, action density) appropriate for dynamics at large distance such as gravitation:
\beq \label{effact}
	\mathcal{L} = \frac{M_{\rm P}^2}{2} \left (R + \partial_\mu \bar{\vp} \partial^\mu \bar{\vp} \right) - M_{\rm P}^4 V(\bar{\varphi}) - V_{\varphi m} + \mathcal{L}_{\rm em} +\cdots 
\eeq
where I have omitted kinetic terms for matter fields, and nuclear interactions. Here $V_{\varphi m}$ contains the interactions of the scalar with particles that constitute ordinary matter, while $\mathcal{L}_{\rm em}$ gives the dependence of the electromagnetic interaction:
\begin{gather}
	V_{\varphi m} = m_{e0} c_e(\bar{\vp}) \bar{e}e + m_{N0} c_N(\bar{\vp}) (\bar{n}n+\bar{p}p) + \frac{1}{2} \delta_{N0} c_v(\bar{\vp}) (\bar{n}n -\bar{p}p) \\
	-\mathcal{L}_{\rm em} = \frac{1}{4}Z(\bar{\vp}) F_{\mu\nu}F^{\mu\nu}, 
\end{gather}	
where $m_{N0}=(m_n+m_p)/2$ and $\delta_{N0}=m_n-m_p$ for present-day values of masses, thus the functions $c_i$ and $Z$ are normalized to unity for the present value $\bar{\vp}_0$. Then
\beq
	\Delta \ln \alpha = - \ln Z(\bar{\vp}), \qquad
	\Delta \ln \mu = \ln \frac{c_N(\bar{\vp}) - \frac{\delta_{N0}}{2m_{N0}}c_v(\bar{\vp}) }
	{1-\frac{\delta_{N0}}{2m_{N0}} c_e(\bar{\vp}) }
\eeq
For a small change in $\bar{\vp}$ about the present value we expand $c_i= 1+\lambda_i(\bar{\vp}-\bar{\vp}_0)+\cdots$ to define the couplings $\lambda_{e,N,v}$ and $Z(\bar{\vp})=(1+\lambda_{\rm em} (\bar{\vp} -\bar{\vp}_0))^{-1}$. \footnote{Under some special assumptions \cite{Damour}, the leading term in the expansion of $\lambda_i$ and $Z^{-1}$ about some particular value may be quadratic in $\bar{\vp}$, then the scalar forces will be much suppressed.} Thus $\Delta \ln \alpha = \lambda_{\rm em}\Delta \bar{\varphi}$ and $\Delta \ln \mu = (\lambda_i-\lambda_v\delta_N/2m_N-\lambda_e) \Delta \bar{\varphi}$. Comparison with Brans-Dicke theory, which may be reformulated as a scalar with universal coupling $(3+2\omega)^{-1/2}$ to matter, indicates that the magnitude of couplings is bounded at a few times $10^{-3}$ before considering differential acceleration. 

We then consider the dependence of atomic mass per baryon $M/A$ on the scalar:
\beq
	\frac{M}{A} = f_p (m_p+m_e) + (1-f_p) m_n + \frac{E_{\rm em}}{A} - 
	\frac{B_{\rm strong}}{A}
\eeq
where the proton fraction $f_p \equiv Z/(Z+A)$. Here $E_{\rm em}$ is the nuclear electromagnetic self-energy, estimated as $(\alpha/\alpha_0) f_{\rm em} m_N Z(Z-1)/A^{1/3}$ with $f_{\rm em}\simeq 7\times 10^{-4}$, and $B_{\rm strong}$ is the nuclear binding energy due to strong interactions. Owing to the complex and poorly known dependence of nuclear forces on fundamental parameters, this last contribution has generally been neglected. To find the differential acceleration of two test bodies towards a given massive source, the source coupling $d \ln (M/AM_{\rm P})_s/d \bar{\vp}$, which gives the value of $\nabla \bar\vp/g$ through solving a Poisson equation for $\bar{\vp}(\vec{r})$, and the test body couplings $d \ln (M/AM_{\rm P})_{1,2}/d \bar{\vp}$, must be evaluated. The source coupling is always dominated by $d\ln c_N/d \bar{\vp}\simeq \lambda_N$, thus we may write 
\beq \label{etalowenergy}
	\eta \simeq \frac{\lambda_N}{2} 
	\left[(-\lambda_v\frac{\delta_N}{m_N} + \lambda_e\mu^{-1}) \Delta_{12}f_p 
	+ \lambda_{\rm em} f_{\rm em} \Delta_{12}\frac{Z^2}{A^{4/3}} \right]
\eeq
where $\Delta_{12}$ denotes the difference between test bodies: typically $\Delta_{12}f_p$ may be of order $0.03$ while $\Delta_{12}(Z^2/A^{4/3})$ is of order 1.\footnote{This treatment is valid provided that the scalar's Compton wavelength (inverse mass) is larger than the distance between source and test bodies. Otherwise we have a suppression factor $e^{-rm_{\bar{\vp}}}$.} 

Within any given unified model, we may define $\lambda_X\equiv d\ln \alpha_X/d\bar{\vp}$ and calculate the low-energy couplings using previous results for nucleon and electron masses and $\alpha$ \cite{me03}: the result for $\eta$ is then proportional to $\lambda_X^2$ \cite{me06}. Next we relate this bound to cosmological and present-day variations of $\alpha$ and $\mu$.

\subsection{Scalar evolution in cosmology and space-time variation}
Clearly by making $\lambda_X$ arbitrarily small any bound from differential accelerations may be evaded; but the cosmological variation also becomes negligibly small, unless the variation of $\bar{\vp}$ is sufficiently large. Here we consider the cosmological effects of an evolving scalar. While the local gravitational effects may be estimated by solving for the scalar profile in a static spacetime, large-scale cosmological effects are found by considering the time evolution of an approximately homogeneous and isotropic metric of Robertson-Walker form $ds^2 = dt^2 - a^2(t)\sum_i dx_i^2$. The equation of motion of the scalar is then 	
\beq
	\ddot{\vp} + 3H \dot{\vp} = - V'(\vp) - V'_{\rm eff}(\vp,\rho_{\rm matter})
\eeq
where $H\equiv \dot{a}/a$ is the Hubble expansion rate. Here $V'_{\rm eff}(\vp,\rho_{m})\sim \rho_m d \ln m/d \bar{\vp}$ arises from the coupling of the scalar to (nonrelativistic) matter, including possible dark matter, $m$ being the mass of the relevant particle. If we assume small couplings, the bare potential $V'(\vp)$ will generally dominate. Note that the effective mass-squared $V''(\vp)$ should be extremely small (comparable with $H^2$) if the value of the scalar is to drift monotonically over cosmological time rather than oscillate about a minimum. The kinetic energy of the field is $T=\dot{\vp}^2/2$, resulting in an effective equation of state $w_\vp \equiv p_\vp/\rho_\vp = (T-V)/(T+V)$. 

The evolution of a scalar may be connected with the most conspicuous apparent deviation from GR on large scales: this is the recent accelerated expansion of the Universe, $\ddot{a}>0$, required to explain a number of independent cosmological observations: see {\em e.g.}\ \cite{Padmanabhan:2002ji}. Acceleration requires a total equation of state $w\leq -1/3$ averaged over all mass-energy components, whereas matter (baryonic or dark) has $w_m\simeq 0$ and radiation has $w_\gamma = 1/3$. Thus some additional form of energy density with negative pressure should be present and dominate over matter and radiation in recent cosmological evolution. This is the so-called ``dark energy''. 

The theoretically simplest candidate for dark energy is a cosmological constant $\Lambda$, which was already allowed by Einstein in the formulation of GR. The corresponding equation of state is $w_\Lambda=-1$, which is compatible with all current observations. However, it is also possible that scalar fields have contributed significantly to the energy density either at present or at earlier epochs. We may either assume that the entire acceleration is due to $\vp$, or use observations to put bounds on the rate of change $\dot{\vp}$ at recent epochs. 

The rate of variation is related to the energy density fraction $\Omega_\vp \equiv \rho_\vp/3H^2M_{\rm P}^2$ via
\beq \label{phidot}
	\frac{\dot{\bar{\vp}}}{H} \equiv \frac{\sqrt{2 T}}{HM_{\rm P}}
	= \sqrt{3 \Omega_\vp (1 + w_\vp)}
\eeq 
where both quantities inside the square root are bounded above observationally. 
\begin{figure}
	\centerline{ 
	\includegraphics[viewport = 120 410 510 580, clip, width=10cm]{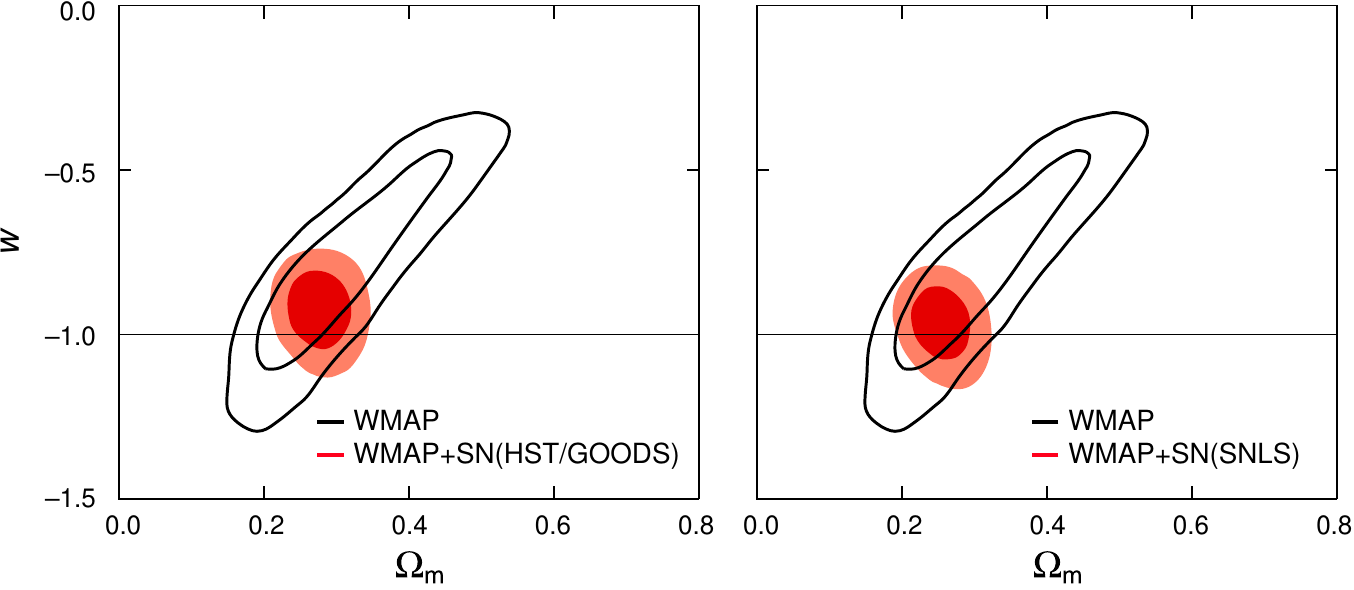}
	} 
\caption {Observational bounds on matter density and dark energy equation of state, from \cite{Spergel:2006hy}. A spatially flat Universe is assumed, for which $\Omega_{\rm de} + \Omega_m \simeq 1.0$. }
\label{spergels}
\end{figure}
Recent measurements give an allowed region in the $\Omega_\vp$--$w_\vp$ plane, Fig.~\ref{spergels}, within which Eq.~(\ref{phidot}) attains a maximum value of approximately $0.7$. Thus the recent scalar evolution cannot have been faster than $\dot{\bar{\vp}}_{\rm max}\simeq 5\times 10^{-11}\,$y$^{-1}$, using $H_0\simeq 7\times 10^{-11}\,$y$^{-1}$. This bound was obtained assuming that $\vp$ accounts for the entire dark energy, but it holds much more widely because 1) the total magnitude of dark energy density $\Omega_{\rm de}$ is the relevant quantity affecting cosmological evolution, and bounds $\Omega_{\vp}$ above; and 2) the scalar kinetic energy influences the total equation of state whether or not $\vp$ constitutes the dark energy. As an extreme example, a scalar mimicking cold dark matter with $w_\vp\simeq 0$ and $\Omega_\vp \simeq 0.3$ would have $\dot{\bar{\vp}}/H\simeq 1$.

We may now write $\partial_t \ln \{ \alpha, \mu \} \leq \{ c_1, c_2 \} \lambda_X \dot{\bar{\vp}}_{\rm max}$ where $c_{1,2}$ are numerical constants that depend on the choice of unified model. For the differential acceleration we have
\beq
	\eta \simeq K\frac{\Delta_{12}f_p}{2} \lambda_X^2
\eeq
where $K$ is another model-dependent constant arising from Eq.~(\ref{etalowenergy}), where we keep only the first term on the RHS which dominates for the models we consider. Thus for a nonzero variation of $\alpha$ or $\mu$, setting $\Delta_{12}f_p=0.036$ we find
\beq
	\eta \leq \frac{K}{c_1^2} \left(\frac{\partial_t \ln \alpha} 
	{3.7\times 10^{-10}\,{\rm y}^{-1}} \right)^2, \qquad 
	\eta \leq \frac{K}{c_2^2} \left(\frac{\partial_t \ln \mu} 
	{3.7\times 10^{-10}\,{\rm y}^{-1}} \right)^2. 
\eeq
For instance, in a scenario where the unified coupling varies but the Higgs v.e.v.\ and superpartner masses do not, $\beta_v=\beta_S=0$ in Eq.~(\ref{betadefin}), we have $c_{1,2} \simeq\{0.52,13\}$ and $K\simeq |\lambda_N\lambda_v\delta_N/m_N| \simeq 0.18$, thus $K/c_{1,2}^2 = \{0.67,1.1\times 10^{-3}\}$. Hence a variation $\dot{\alpha}/\alpha\simeq 10^{-16}$ per year would lead to $\eta \leq 5\times 10^{-14}$. E{\" o}tv{\" o}s experiments may be competitive with atomic clocks in bounding variations in such models \cite{me06,Wettericheta}. 

Conversely, in a scenario where where $\beta_v\simeq \beta_S\simeq 34$, we find $c_{1,2}\simeq \{1.0,4\}$ and $K \simeq 1.5$, thus $K/c_{1,2}^2 = \{1.5,9.5\times 10^{-2}\}$. The small variation of $\mu$ is due to a cancellation between variations of $m_p/M_X$ and $m_e/M_X$, whereas the scalar couplings to the proton and electron remain relatively large. Thus for a linear time variation $\dot{\mu}/\mu \simeq 2\times 10^{-15}\,$y$^{-1}$, which would yield $\Delta\mu/\mu \simeq 2\times 10^{-5}$ at high redshift, this scenario predicts $\eta>10^{-12}$, already beyond experimental bounds. 

Several of the assumptions we made in deriving these results may be violated: for example scalars may be significantly heavier than the Hubble scale, or may have couplings to matter or self-couplings that are not weak. The effective scalar mass may well depend on $\rho_m$, which is the special feature of so-called ``chameleon'' models \cite{chameleon} whose gravitational effects are quite different from very light, weakly-coupled fields. In such cases the dominant variation of the scalar is likely to be with respect to spatial position or local density \cite{EllisOlive} rather than a slow cosmological drift. The framework of an effective action for the scalar and matter fields may also be used in these cases to predict the astrophysical or cosmological behaviour of such theories.


\section{Conclusion}
I have sketched the theoretical framework of the Standard Model and General Relativity necessary to understand the r{\^ o}le of ``fundamental constants'' in theoretical physics, and discussed theoretical extensions which could account for the values of these parameters and their possible cosmological variation. There are many potential and actual observational probes of such new physics associated with ``varying constants''. Determining their history brings together many branches: astrophysics, optics, gravitation, and atomic, nuclear and particle physics. At present we are still awaiting a clear signal of non-standard behaviour; nevertheless the search for variation, and the associated violation of equivalence principles, involves some of the most sensitive and precise experiments in physics. 

\subsection*{Acknowledgements}
I would like to thank Steffen Stern and Christof Wetterich for their collaboration and insights on the subjects covered by this review, and Iain Brown and Felix Br{\" u}mmer for discussions while preparing the manuscript. The author is supported by by the {\it Impuls- and Vernetzungsfond der Helmholtz-Gesellschaft}.

%

\end{document}